\documentclass[useAMS,usenatbib]{mn2e}

\usepackage{amsmath,array,amsfonts}
\usepackage{amssymb}
\usepackage{graphicx}
\usepackage{natbib}
\usepackage{pstricks}
\usepackage{verbatim}
\usepackage{multirow}
\usepackage{bbold}

\newcommand{\ltsima} {$\; \buildrel < \over \sim \;$}
\newcommand{\gtsima} {$\; \buildrel > \over \sim \;$}
\newcommand{\lta} {\lower.5ex\hbox{\ltsima}}
\newcommand{\gta} {\lower.5ex\hbox{\gtsima}}

\title[On the void explanation of the Cold Spot]{On the void explanation of the Cold Spot}

\author[A. Marcos-Caballero et al.]{A. Marcos-Caballero$^1$$^,$$^2$, R. Fern\'andez-Cobos$^1$,
E. Mart\'\i nez-Gonz\'alez$^1$, P. Vielva$^1$\\
$^1$     Instituto de F\'isica de Cantabria, CSIC-Universidad de Cantabria, Avda. de los Castros s/n, E-39005 Santander, Spain.\\
$^2$     Departamento de F\'isica Moderna, Universidad de Cantabria, Avda. los Castros s/n, E-39005 Santander, Spain.}

\date{Accepted  Received ; in original form }

\begin{document}

\maketitle

\begin{abstract}
The integrated Sachs-Wolfe (ISW) contribution induced on the cosmic
microwave background by the presence of a supervoid as the one
detected by \citet{Szapudi2015} is reviewed in this letter in order to
check whether it could explain the Cold Spot (CS) anomaly. Two
different models, previously used for the same purpose, are considered
to describe the matter density profile of the void: a top hat function
and a compensated profile produced by a Gaussian potential. The
analysis shows that, even enabling ellipticity changes or different
values for the dark-energy equation of state parameter $\omega$, the
ISW contribution due to the presence of the void does not reproduce
the properties of the CS.
\end{abstract}
\section{Introduction}
\label{sec:Introduction}
The Cold Spot (CS), an extremely cold region centred on $(b, \ell) =
(210^{\mathrm{\circ}},-57^{\mathrm{\circ}})$, was discovered in the
Wilkinson Microwave Anisotropy Probe (WMAP) data using a multiscale
analysis of the Spherical Mexican Hat Wavelet (SMHW) coefficients
\citep{Vielva2004, Cruz2005}. Within the $\mathrm{\Lambda CDM}$ model,
the significance of the occurrence of this feature in the cosmic
microwave background (CMB) anisotropies was estimated between $1\%$
and $2\%$ \citep{Cruz2006}. As the Planck Collaboration confirmed, the
CS shows unusual properties which come to light when the mean angular
profile or the area of wavelet coefficients above a certain threshold
on angular scales around $10^{\mathrm{\circ}}$ are analysed
\citep{Planck2015XVI}. Besides the possibility that the CS could
  be a statistical fluke, different explanations have been proposed.
Although this letter is focused on the void hypothesis, other physical
mechanisms include a cosmic bubble collision \citep{Czech2010,
  Feeney2011, McEwen2012}, the gravitational evolution of a cosmic
texture \citep{Cruz2007}, and alternative inflationary models
\citep{BuenoSanchez2014}.

Recently, there has been a debate on whether the CS could be explained
as a consequence of the presence of a large void, which was detected
in the WISE-2MASS galaxy survey at the same direction
\citep{Szapudi2015, Finelli2014}. Actually, this is not the first time
in which a void arises as the possible origin of the CS \citep[see
  e.g.][]{Tomita2005, Inoue2006, Rudnick2007, Cruz2008, Bremer2010,
  Granett2010}. This low-density region is consistent with a supervoid
centred at $z\approx 0.15 - 0.25$, depending on its
characterization. The alignment of the void and the CS is pointed out
as a hint of a physical connection between both phenomena. They built
their argument based on a probabilistic discussion about this
alignment and a particular case of the Lema\^itre-Tolman-Bondi
  (LTB) model with a Gaussian potential \citep{Finelli2014} to infer
the angular profile of the CMB imprint of a spherically symmetric
supervoid in the number density of galaxies. In this latter paper, the
connection between the supervoid detected in WISE-2MASS and the CS was
analysed in the light of the integrated Sachs-Wolfe (ISW) and the
Rees-Sciama contributions. However, \citet{Zibin2014} and
\citet{Nadathur2014} show independently that the first-order ISW
contribution from the presence of this type of void is actually
dominant with respect to the non-linear component (Rees-Sciama
effect), and therefore the corresponding temperature decrement induced
in the CMB by the presence of a void as the one mentioned above
($\approx -19\mathrm{\mu K}$) would not be intense enough to account
for the depth of the CS ($\approx -150 \mathrm{\mu K}$).

In this letter, we explore the latter argument through a supplementary
analysis in the SMHW coefficients \citep{Martinez2002} at the specific
CS angular scale, since the anomaly is detected in the SMHW space. In
addition, we extend the void models enabling ellipticity changes to
check that a different geometry could not produce an ISW contribution
which accounts for the CS. We also show that alternative simple models
of dark energy cannot reconcile the CMB contribution from a supervoid
and the observed CS temperature. Finally, we discuss the previous
analyses.

%
 
\section{The void influence on the CMB}
\label{sec:isw}
As it is known, within the standard cosmological model, the
contribution of any possible supervoid is already included in the
total CMB anisotropies (as a part of the linear ISW contribution) and
therefore the presence of a standard and linear underdensity cannot
explain the anomalous temperature decrement of the CS. The assumption
that the effect on the CMB photons due to the nonlinear evolution of
the potential is negligible with respect to the ISW contribution is
based on previous analyses of the Rees-Sciama contribution, which
becomes noticeable at multipoles $\ell > 80$ ($\lesssim 2^{\circ}$),
and even at these angular scales, its value is much lower than the ISW
component at large scales \citep[see e.g.][]{Cai2010}. Therefore, a
rare void is needed in order to explain the CS with the ISW and
Rees-Sciama effects. These non-standard scenarios are explored varying
the void eccentricity up to very unlikely values. In any case, the
angular size of the ISW effect of the voids considered in this work is
greater than several degrees.

Besides the amplitude of this decrement, the profile of the CS
is also important to characterize the anomaly because a particular
shape is preferred when it is selected in the SMHW coefficients. In
this section, we first review the main conclusions about the ISW
contribution expected from the presence of a void as that detected by
\citet{Szapudi2015}. Subsequently, the impact of varying the
ellipticity of the void is also explored. In addition, non-standard
scenarios with different values of $\omega$ are considered to check
whether the void prediction is able to cause a temperature decrement
as that observed in the CS.

\subsection{Spherical model}
Because of symmetry assumptions, the ISW contribution to the CMB anisotropies caused by a large-scale structure (LSS) fluctuation can be written as:
\begin{equation}
\small
\dfrac{\Delta T (\theta)}{T_{\mathrm{CMB}}} = -2\int {\mathrm{d}z \dfrac{\mathrm{d}G(z)}{\mathrm{d}z}\Phi\left( \sqrt{\chi^2(z)+\chi_0^2-2\chi(z)\chi_0\cos\theta}\right)},
\label{eq:profile}
\end{equation}
where $\theta$ denotes the angular distance from the centre of the void at $\chi_0 = \chi(z_0)$, in comoving distance. The gravitational potential $\Phi(\mathbf{r},z)$ is factorized into the growth suppression factor $G(z)$ and a spatial dependence $\Phi(r)$ which, assuming $G(0)=1$, represents the potential at $z=0$. 

In this letter, two different density profiles, which have been already used to the same purpose, are considered. On the one hand, a spherical top hat (TH) model \citep{Szapudi2015}, parametrized by its radius $R$. In this case, the potential can be written as
\begin{equation}
\Phi(r) =\begin{cases} \phi_0 R^2 \left( 3- \dfrac{r^2}{R^2}\right),  & \mbox{if } r\le R \\
                       \phi_0 \dfrac{2R^3}{r},  & \mbox{if } r> R, \end{cases}
\end{equation}
where $r$ denotes the comoving distance from the centre of the void.

When distances greater than $R$ are considered, this model behaves as a point-like particle: it presents an inverse dependence on distance, and therefore the gravitational effect is extended far beyond distances as the size of the void.

On the other hand, a particular case of LTB model is considered
\citep{Finelli2014, Nadathur2014}. The potential is described in this
case by a Gaussian profile:
\begin{equation}
\Phi(r) = \phi_0 {r_0}^2\exp{\left(-\dfrac{r^2}{{r_0}^2}\right)},
\end{equation}
where $r_0$ accounts for the scale. Hereafter, this profile is
referred as the Gaussian model, although the matter underdensity
profile is not Gaussian in this case\footnote{Notice that this model
  is denoted simply as LTB in previous papers
  \citep{Szapudi2015,Finelli2014,Nadathur2014}.}.

It is easy to show that, whilst the density profile associated to the
Gaussian potential is compensated, that associated to the TH
model is not.

In both cases, the amplitude $\phi_0$ is proportional to the matter density fluctuation at the void centre $\delta_0$:
\begin{equation}
\phi_0 = \dfrac{\Omega_m \delta_0}{4G(0)}\left( \dfrac{H_0}{c}\right)^2,
\end{equation}
where, in a flat universe, $\Omega_m = 1 - \Omega_{\Lambda}$ denotes
the matter energy density (in our case, with a fixed dark-energy
density $\Omega_{\Lambda} = 0.685$), $H_0$ is the Hubble constant at
present time and $c$ is the speed of light in vacuum.

The best-fitting set of parameters is considered for each model. In
particular, we take $R = (220\pm50) \mathrm{h^{-1}Mpc}$, $\delta_0 =
0.14\pm0.04$ and $z_0 = 0.22\pm 0.03$, for the TH model
\citep{Szapudi2015}; and $r_0 = (195\pm35) \mathrm{h^{-1}Mpc}$,
$\delta_0 = 0.25\pm0.10$ and $z_0 = 0.155\pm 0.037$, in the case of
the LTB Gaussian model \citep{Finelli2014, Nadathur2014}.

In order to characterize the feature induced in the CMB temperature anisotropies by the presence of a supervoid, we compute its 1-dimensional shape. This profile can be expanded in terms of the Legendre polynomials:
\begin{equation}
\dfrac{\Delta T(\theta)}{T_{\mathrm{CMB}}} = \sum_{\ell = 0}^{\infty}{\sqrt{\dfrac{2\ell+1}{4\pi}} a_{\ell} P_{\ell}(\cos\theta)},
\end{equation}
where $a_{\ell}$ denotes the coefficients of the expansion. In the particular case in which the void is aligned with the z-axis, the coefficients $a_{\ell}$ are equivalent to the spherical harmonic coefficients with $m=0$. They can be therefore computed from the theoretical profile of Eq. (\ref{eq:profile}) as
\begin{equation}
a_{\ell} = \sqrt{\left( 2\ell + 1\right)\pi}\int_{-1}^{1}{\mathrm{d}(\cos\theta)\dfrac{\Delta T(\theta)}{T_{\mathrm{CMB}}}P_{\ell}(\cos\theta)}.
\end{equation}

\begin{figure}
\centering
\includegraphics[scale=0.32,angle=270]{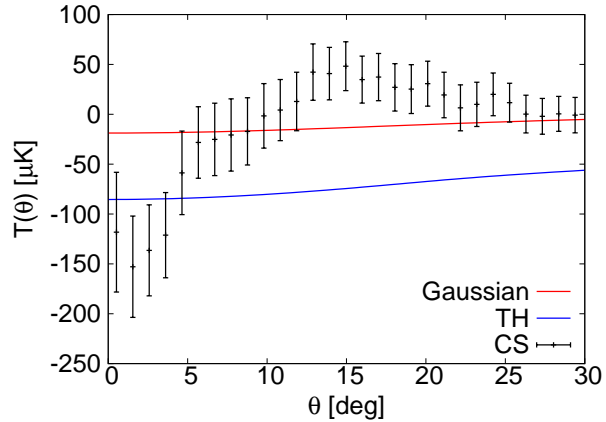}
\caption{CMB temperature profiles induced by the presence of a
  supervoid modelled as a TH (in blue) and a Gaussian model (in
  red). The data points correspond to the CS profile from the
    Planck SMICA map, and the error bars represent the cosmic
    variance.}
\label{fig:profiles}
\end{figure}

The corresponding ISW profiles induced by each void model {and the CS
  data} are depicted in Figure~\ref{fig:profiles}. The profiles are
very different in terms of the amplitude. Within the considered
$\Lambda\mathrm{CDM}$ model, the standard deviation of the ISW
  temperature fluctuations is estimated to be $\sigma_{\mathrm{ISW}}
= 19.58\mathrm{\mu K}$. Whilst the Gaussian model induces a profile
whose value at $\theta = 0$ lies at the $1\sigma$ level when the
standard deviation due exclusively to the ISW contribution is taken as
reference, the TH profile at the centre reaches a $4.5\sigma$ level.

In terms of the standard deviation of the matter field convolved by a
top hat function of scale $R$, the corresponding value of $\delta_0$
for the TH best-fit profile lies at the $\approx 6\sigma$
level\footnote{Notice that \citet{Szapudi2015} provide a value of at
  least $3.3\sigma$ based on a more conservative estimate of the
  rareness of the void which takes into account a $1\sigma$ deviation
  of the TH best-fit parameters.}. This could give a hint that the TH
model is not a realistic description of a void expected within the
standard model, although it is shown closer --but not enough yet-- to
explain the CS anomaly. Actually, this void description would imply an
anomaly larger than the one that is expected to be explained. For the
Gaussian model, the value of $\delta_0$ is only at a
$\approx2\sigma$ level.


In addition to the amplitude, a deeper insight can be obtained by
paying attention to the shape of the profile. The SMHW
  coefficient of the CS with scale $R=300 \arcmin$ describes both the
  temperature at the centre and the hot ring at $15^\circ$, since the
  specific shape of the SMHW at this scale weighs these features in a
  single number. Therefore, if the theoretical profiles fit the CS
  data, they will have a similar value of the SMHW coefficient. It is
  also important to remark that the CS represents a $\approx
  4.7\sigma$ fluctuation in terms of this coefficient, which implies
  that any theoretical model assumed for the CS must explain this
  large deviation. The value of the SMHW coefficient can be computed
  as
\begin{equation}
W_0 = \sum_{\ell=0}^{\infty}{\sqrt{\dfrac{2\ell+1}{4\pi}}w_{\ell}a_{\ell}}.
\end{equation}

The standard deviation of the SMHW coefficients with $R=300 \arcmin$
(the scale at which the CS anomaly is manifested) due to the ISW
contribution is $\sigma_{\mathrm{ISW}}(W_0) = 0.94\mathrm{\mu K}$. We
obtain $W_0$ values at around $-1.07 \mathrm{\mu K}$ for the TH
description and $-0.54 \mathrm{\mu K}$ for the Gaussian model, and
both lie within the $\approx 1\sigma$ level when only the ISW
contribution is taken into account. On the other hand, the SMHW
coefficient associated to the CS is a $20\sigma$ fluctuation with
respect to the ISW effect, and therefore is very unlikely to explain
the CS only taking into account the ISW fluctuations of linear
standard voids. Other possible scenario is that the CS is the sum of a
primordial CMB fluctuation and the ISW effect of a void, but even in
this case the probability of this event is small. The SMHW coefficient
of the observed data, once the effect of the void is subtracted, is
still a $\approx 4.5 \sigma$ fluctuation. Therefore, whilst the effect
predicted by the theoretical models for this particular void is shown
compatible with the expected ISW signal from typical LSS fluctuations
within the $\Lambda \mathrm{CDM}$, the CS appears anomalous in
relation to both properties: shape and amplitude.

In principle, to consider the void as explanation of the CS, it would
not be necessary that its contribution accounts for all the CS
amplitude, but it should be intense enough to make anomalous the
primordial fluctuation. In terms of the amplitude of the Gaussian
model, the ISW contribution from the void represents a $13\%$ with
respect to the temperature at the centre of the CS. However, in terms
of $W_0$, this fraction drops to $2.8\%$.

\subsection{Ellipsoidal model}
All previous conclusions are derived from a spherical void model, but we could wonder whether they remain when the void presents an ellipsoidal geometry. For this purpose, we decompose the radial coordinate $\mathbf{r}$ of the matter density profile, defined from the centre of the void, into a component parallel to the line of sight $r_{\parallel}$ and another orthogonal to it $\mathbf{r_{\perp}}$, which is a 2-dimensional vector in the normal plane, such that:
\begin{equation}
r = \sqrt{r_{\parallel}^2 \left( 1-e^2\right)+r_{\perp}^2},
\end{equation}
where $e$ denotes the ellipticity. This toy model allows us to stretch
the void along the line of sight in terms of the ellipticity, whereas
the semi-minor axis is fixed to the scale of the density profile ($R$
for the TH and $r_0$ for the Gaussian model, respectively), implying
an increase of the volume. The centre position of the void is also
kept at $z_0$. This configuration favours the increase of the ISW
contribution due to the presence of the void, because the void
influence is kept in a greater redshift interval along the line of
sight.

\begin{figure}
\centering
\includegraphics[scale=0.32,angle=270]{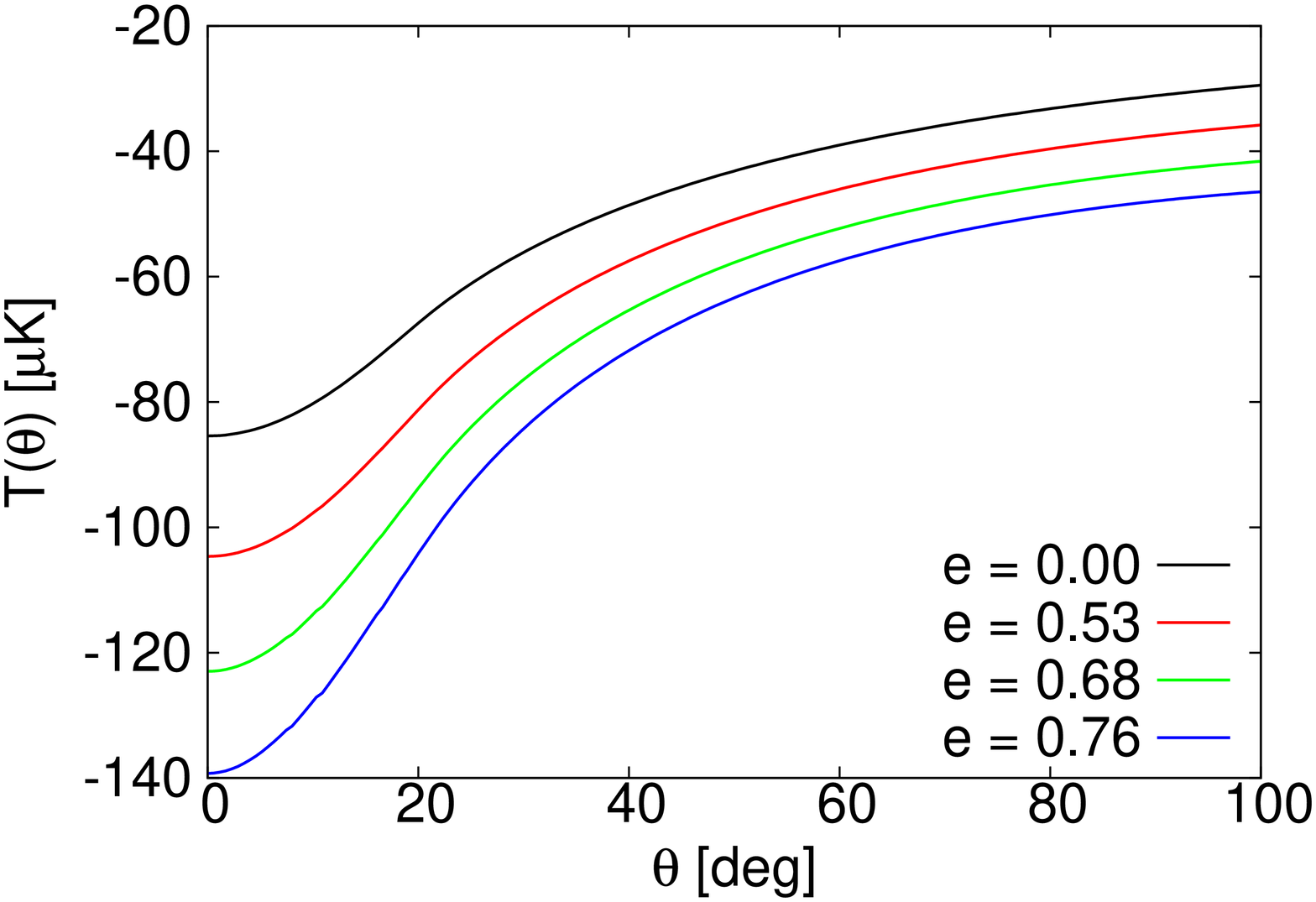}
\includegraphics[scale=0.32,angle=270]{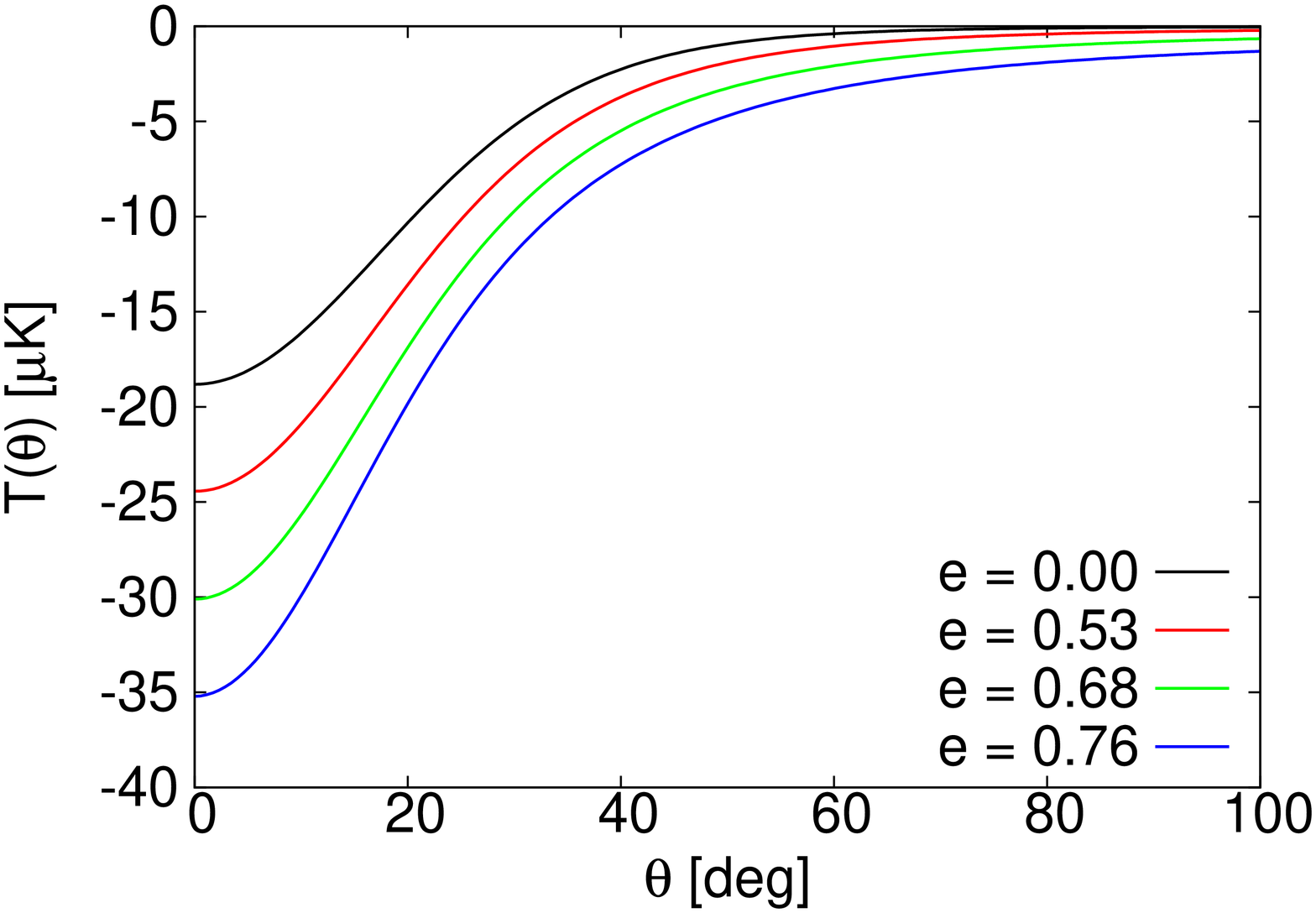}
\caption{Comparison of CMB temperature profiles induced by the
  presence of an elliptical supervoid modelled as a TH (top panel)
  and a Gaussian model (bottom panel) with different values of
  ellipticity.}
\label{fig:profiles_e}
\end{figure}

Although the standard model imposes limits to the ellipticity
\citep[e.g.][]{Icke1984, Bardeen1986}, three values are considered
such that the semi-major axis is increased by one, two and three times
the error bar of $r_0$ (the value of $35\mathrm{h^{-1}Mpc}$ is taken
in both models for simplicity). A comparison between CMB temperature
profiles caused by supervoids with different ellipticity is shown in
Figure \ref{fig:profiles_e}. As expected, the absolute value of the
amplitude at $\theta = 0$ increases as the ellipticity grows. In the
case of the TH model, the radial profile at the centre of the void
reaches a value close to the CS temperature decrease when an
ellipticity of $e=0.76$ is considered, whilst these values remain
unreachable with the Gaussian model. However, all the SMHW
coefficients lie within the $1\sigma$ level of the ISW contribution,
as in the spherical case. This means that the shape of the profiles
differs from that shown by the CS. The $W_0$ value for all cases are
given in Table \ref{tab:coef_e}. They should be compared with the SMHW
coefficient at the CS location in the \textit{Planck} temperature data
whose value is estimated in $-19.3 \pm 4.1 \mathrm{\mu K}$.

\begin{table}
\centering
\begin{tabular}{ccc}
\hline \hline $e$ & TH [$\mathrm{\mu K}$] & Gaussian
       [$\mathrm{\mu K}$] \\ \hline 0.00 & -1.07 & -0.54 \\ 0.53 &
       -1.42 & -0.71 \\ 0.68 & -1.81 & -0.85 \\ 0.76 & -2.20 & -1.03
       \\ \hline \hline
\end{tabular}
\caption{SMHW coefficients $W_0$ induced by elliptical voids modelled
  by TH and Gaussian profiles with different
  ellipticity. All coefficients correspond to a wavelet scale
  $R=300\arcmin$. The $W_0$ computed at the CS location in the
  \textit{Planck} temperature data is $-19.3 \pm 4.1 \mathrm{\mu K}$.}
\label{tab:coef_e}
\end{table}

\subsection{Varying $\omega$ in the dark-energy equation of state}
\label{sec:beyond}
Assuming $\Lambda\mathrm{CDM}$, $\Omega_{\Lambda}$ regulates the
amplitude of the ISW effect produced by these void models. Considering
dark energy, the ISW contribution also depends on its evolution. In
this section, we extend the void models so that the dark-energy
equation of state parameter $\omega$ can be set to another value
different from $-1$. This dependence affects explicitly to the growth
suppression factor $G(z)$ and the comoving distance
$\chi(z)$. Decreasing the value of $\omega$ causes a stronger
evolution in the density parameter of the dark energy, implying a
larger ISW imprint. Actually, for our purposes, the assumption that
the $\omega$ is different from $-1$ is only necessary at the redshift
interval in which the CMB photon is suffering the effect of the void
but not in the whole evolution of the Universe.

\begin{figure}
\centering
\includegraphics[scale=0.32,angle=270]{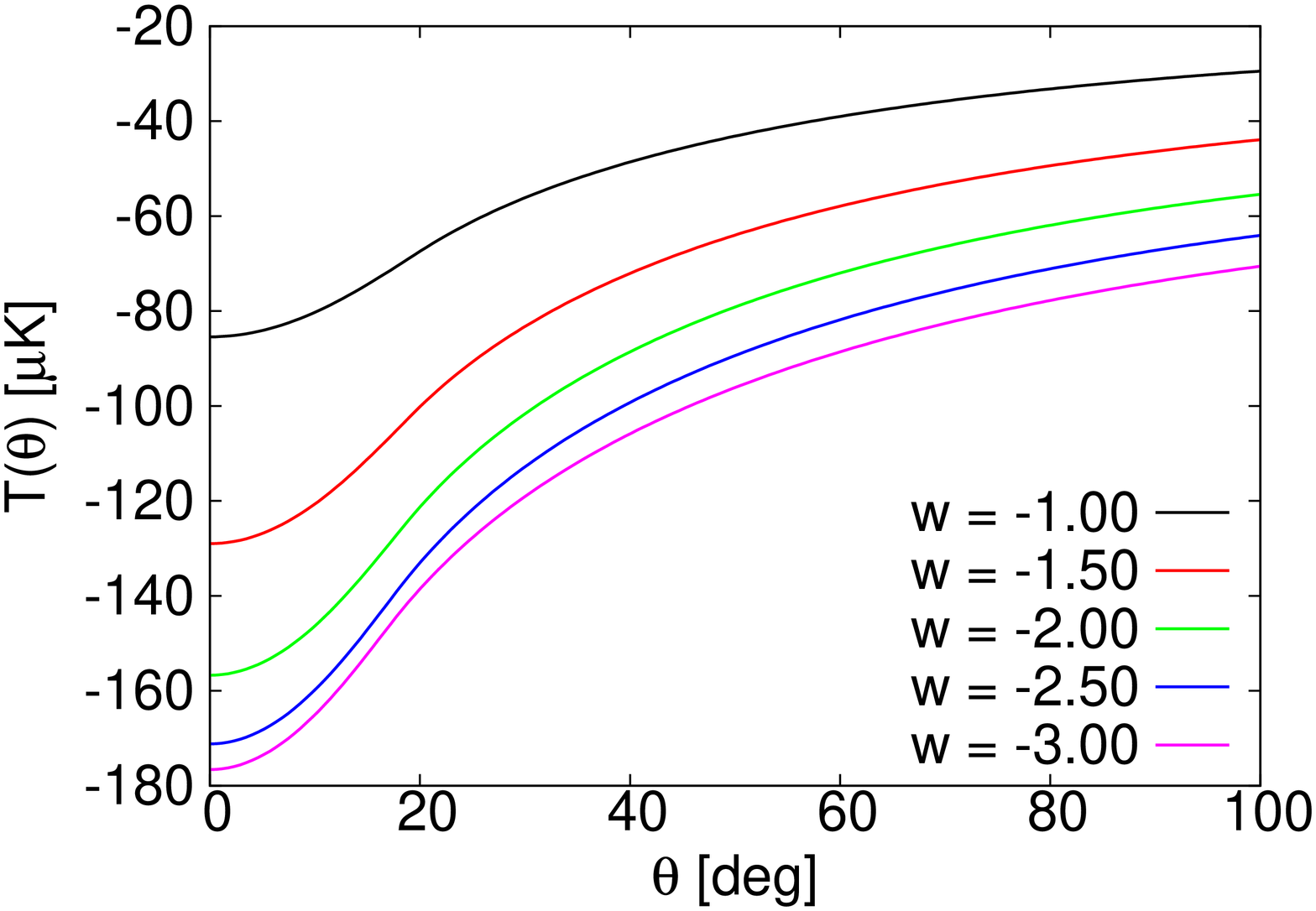}
\includegraphics[scale=0.32,angle=270]{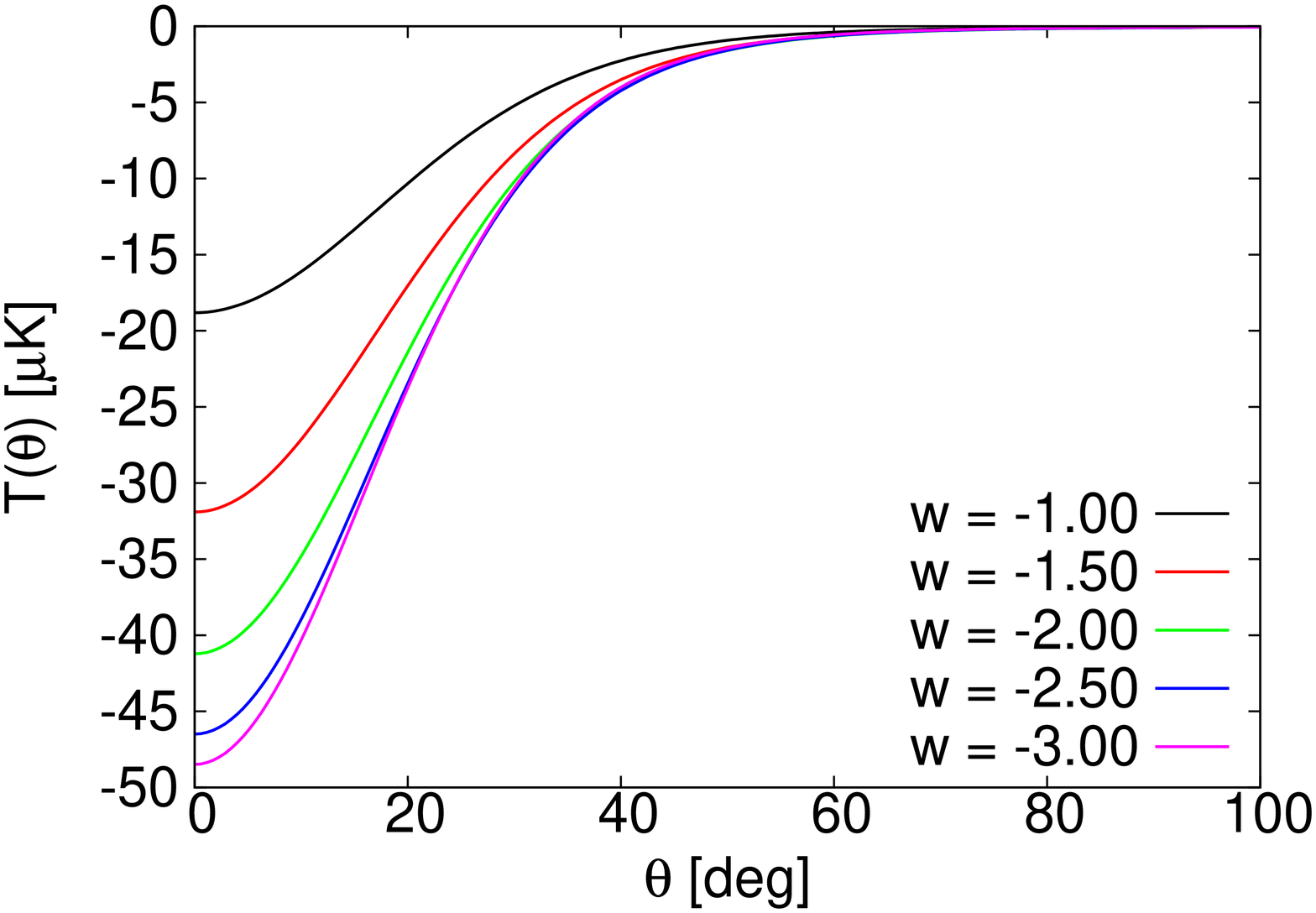}
\caption{Comparison of CMB temperature profiles induced by the
  presence of a spherical supervoid modelled as a TH (top panel) and
  a Gaussian model (bottom panel) with different values of
  $\omega$.}
\label{fig:profiles_w}
\end{figure}

A comparison between CMB temperature profiles induced by the void
corresponding to different values of $\omega$ is given in Figure
\ref{fig:profiles_w}. The temperature at the centre reaches a similar
value than that shown by the CS only for the TH model and considering
a value of $\omega = -3.0$ which, obviously, is ruled out by current
observations \citep[e.g.][]{Parameters2015}. Similar intervals in
$\omega$ does not correspond with similar increases of the absolute
value of the amplitude of the profiles, but this increase is smaller
as the values of $\omega$ become more extreme. However, the $W_0$
values for these profiles also lie within the $1\sigma$ level with
respect to the standard deviation of the ISW signal. They are shown in
Table \ref{tab:coef_w}.

\begin{table}
\centering
\begin{tabular}{ccc}
\hline \hline $\omega$ & TH [$\mathrm{\mu K}$] & Gaussian
       [$\mathrm{\mu K}$] \\ \hline -1.00 & -1.07 & -0.54 \\ -1.50 &
       -1.74 & -0.96 \\ -2.00 & -2.13 & -1.28 \\ -2.50 & -2.34 & -1.49
       \\ -3.00 & -2.38 & -1.60 \\ \hline \hline
\end{tabular}
\caption{SMHW coefficients $W_0$ induced by a spherical void as that
  detected by \citet{Szapudi2015} modelled by TH and Gaussian
  profiles for different values of $\omega$. All coefficients
  correspond to a wavelet scale $R=300\arcmin$. The $W_0$ computed at
  the CS location in the \textit{Planck} temperature data is
  $-19.3 \pm 4.1 \mathrm{\mu K}$.}
\label{tab:coef_w}
\end{table}

\section[Discussion]{Discussion}
\label{sec:Conclusions}
We have reviewed the ISW contribution from a supervoid as the one
detected by \citet{Szapudi2015} in the light of two different models
previously considered: a TH matter density profile and a particular
case of the LTB model with a Gaussian potential. The comparison
between the feature induced on the CMB by the presence of a void as
the one mentioned above and the CS has been focused both on the
amplitude of the induced CMB temperature decrement and the shape of
the radial profile. This is an important aspect, which is related to
the anomalous nature of the CS that is manifested when the CMB is
analysed in wavelet space. As was mentioned in \citet{Planck2015XVI},
the shape of the CS radial profile is shown anomalous, and therefore
the ability to relate this shape with the imprint of a supervoid would
give weight to the hypothesis that there is a connection between both
phenomena. However, an SMHW coefficient analysis shows that the imprint
of the void does not fit the same pattern than the CS profile. All
SMHW coefficients computed in this work lie within the $2.5\sigma$ level
with respect to the standard deviation due to the ISW signal, even for
extreme scenarios that, although discarded within the standard
cosmological model, could provide CMB decrements at the centre of the
CS of the order of the observed data. In the light of these models, it
is important to recall that the ISW imprint from an individual void is
indistinguishable from the primordial fluctuations.

Modifications of the LTB density profile have been considered to
describe more accurately the particular shape of the CMB profile
around the CS \citep[see][]{Finelli2014}. However, the shape is
modified at the expense of a lower value of the amplitude at the
centre, and therefore this amplitude is not already significant. In
fact, we have checked that the $W_0$ values associated with this
profile are even smaller than those related to the cases considered in
this work.

In conclusion, we have shown that the ISW effect within the standard
model is not a plausible explanation for the CS, not even considering
the Rees-Sciama effect. Nevertheless, any hypothetical physical
connection between the void and the CS should rely either on
deviations from the standard cosmological model (e.g. non-Gaussian
primordial density fluctuations) or on new physics.



\section*{acknowledgements}
The authors thank J. Zibin for his useful comments on the
letter. Partial financial support from the Spanish Ministerio de
Econom\'{i}a y Competitividad Projects AYA2012-39475-C02-01 and
Consolider-Ingenio 2010 CSD2010-00064 is acknowledged.
\bibliographystyle{mn2e}
\bibliography{citas_letter}

\end{document}